# COL-CON: A VIRTUAL REALITY SIMULATION TESTBED FOR EXPLORING COLLABORATIVE BEHAVIORS IN CONSTRUCTION

Liuchuan Yu[1,*], Ching-Yu Cheng[2], William F Ranc[1], Joshua Dow[1], Michael Szilagyi[1], Haikun Huang[1], Sungsoo Ray Hong[1], Behzad Esmaeili[2], Lap-Fai Yu[1]

[1] **George Mason University**
[2] **Purdue University**

[*] **lyu20@gmu.edu**

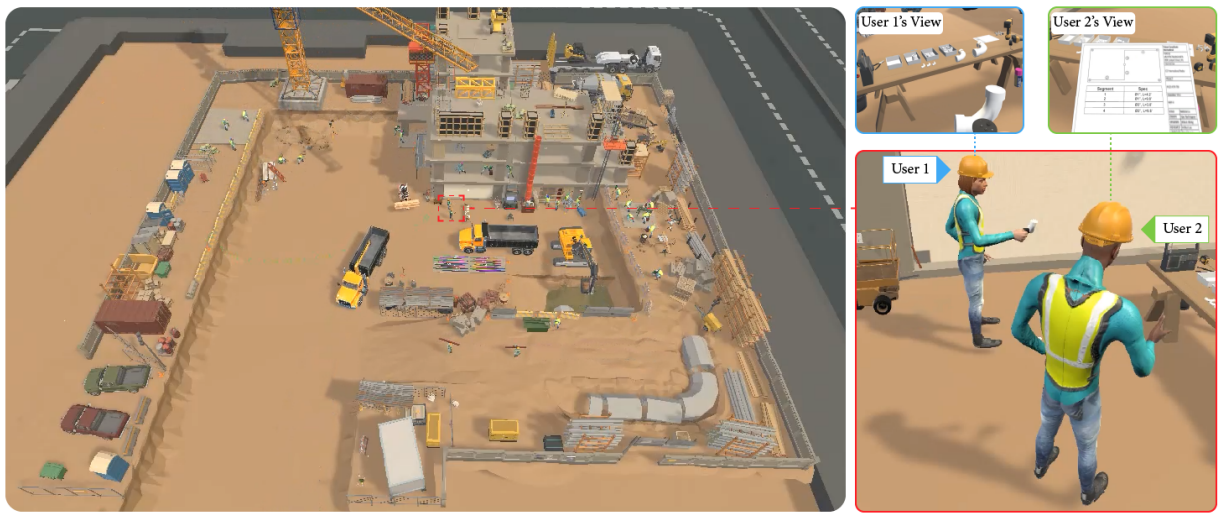

A screenshot of Col-Con. Construction vehicles, including a tower crane, a crane truck, an excavator, a truck, and a forklift, are in operation. Non-player characters are actively engaged in various tasks. Two users can join this immersive testbed to complete tasks collaboratively. We have implemented a realistic pipe installation task where two users (User 1 and User 2) work together to install pipes.

## ABSTRACT

Virtual reality is widely adopted for applications such as training, education, and collaboration. The construction industry, known for its complex projects and numerous personnel involved, relies heavily on effective collaboration. Setting up a real-world construction site for experiments can be expensive and time-consuming, whereas conducting experiments in VR is relatively low-cost, scalable, and efficient. We propose Col-Con, a virtual reality simulation testbed for exploring collaborative behaviors in construction. Col-Con is a multi-user testbed that supports users in completing tasks collaboratively. Additionally, Col-Con provides immersive and realistic simulated construction scenes, where real-time voice communication, along with synchronized transformations, animations, sounds, and interactions, enhances the collaborative experience. As a showcase, we implemented a pipe installation construction task based on Col-Con. A user study demonstrated that Col-Con excels in usability, and participants reported a strong sense of immersion and collaboration. We envision that Col-Con will facilitate research on exploring virtual reality-based collaborative behaviors in construction.

# 1  Introduction

Virtual reality (VR) is increasingly utilized as a mechanism for training Kaplan et al. [2021]. While setting up a real-world training environment can be expensive and time-consuming Karabiyik et al. [2019], training in VR is relatively low-cost, scalable, and effective in improving human performance Koutitas et al. [2020]. Moreover, VR has been applied to the construction industry for various purposes, including workspace planning Getuli et al. [2020], safety training Zhang et al. [2022], Li et al. [2022], Rokooei et al. [2023], Isingizwe et al., and building inspection Albeaino et al. [2022].

Furthermore, construction workers typically collaborate toward a common goal, making effective collaboration essential. Since setting up a real construction site for research purposes can be expensive and risky, there is a need to develop a construction simulation where workers can virtually work together. Most VR applications in construction are still based on single-user experiences, despite the fact that a construction site is a complex environment where various workers, vehicles, tasks, and events coexist. Although context can significantly impact human behavior and performance, these VR applications often focus on single tasks without considering the surrounding environment.

To address these gaps, we propose Col-Con, a virtual reality simulation testbed for exploring collaborative behaviors in construction. Col-Con is designed in a modular fashion, consisting of several modules: the basic scene, vehicles, events, scenarios, tasks, and interactions. We use YAML-based files to facilitate easy simulation setup. Each vehicle is associated with events; each scenario consists of events; and a simulation includes scenarios and tasks. A customized timeline-based event system drives the simulation. Participants can join the simulation from different physical locations to collaborate on tasks. During the simulation, transformations, animations, sounds, and interactions are synchronized, ensuring all participants share the same immersive experience. Moreover, real-time voice communication is embedded to enhance collaboration.

Col-Con differentiates itself from previous VR construction platforms in several key aspects. First, Col-Con is designed as a versatile testbed that is not limited to specific construction tasks. Second, Col-Con offers subjects and researchers an immersive and comprehensive VR construction site where construction vehicles, events, and scenarios can be easily configured. Third, Col-Con is a multi-user simulation testbed that supports the exploration of collaborative behaviors in a VR environment. The transformations, animations, and sounds of the surroundings, users's movements, and interactions are all synchronized. Benefiting from its modular architecture, the construction tasks are independent and can be extended to support additional tasks.

We implemented a pipe-installing task on Col-Con, where two participants collaborate to install pipes on the wall. In addition to human-human collaboration, we explored futuristic construction scenarios involving human-robot and human-AI interactions in the pipe-installation task. Examples include commanding a drone to deliver pipes, directing a robotic dog to process pipes or deliver connectors, and instructing a non-player character (NPC), acting as the experimenter, to perform refilling operations. We conducted a user study to evaluate usability, immersiveness, motion sickness, and task cohesion. Our evaluation results indicate that Col-Con is highly usable, immersive, and suitable for collaborative research.

The major contribution of this work includes:

- Devising a synchronized virtual reality simulation tested for exploring collaborative behaviors in construction called Col-Con. It consists of a realistic construction scene, various construction vehicles and their associated events, and a customized timeline-based event system. It is highly configurable and pluggable based on configuration files, enabling researchers to easily build simulations.
- Implementing a functional construction task on Col-Con. Procedural model generation, arbitrary pipe connecting, rich interactions, and compensation algorithms are applied to realistically simulate pipe-installing tasks. Two users are assigned different roles—the Installer and the Fetcher—to complete tasks collaboratively. Furthermore, human-robot and human-AI interactions are implemented to explore futuristic construction scenarios.
- Conducting a user study involving 14 groups (28 participants) to evaluate Col-Con. The results indicate that Col-Con demonstrates strong usability, provides an immersive experience, has acceptable levels of motion sickness, and fosters a strong sense of collaboration among participants. We also discussed collaborative behaviors on Col-Con.

## 2 Related Work

### 2.1 Extended Reality for Construction

Extended reality (XR) has been widely applied in the Architecture, Engineering, and Construction (AEC) industry Dunston and Wang [2005], Alizadehsalehi et al. [2020], Delgado et al. [2020], Ververidis et al. [2022], Safikhani et al. [2022], Mitterberger [2022], Trindade et al. [2023]. For example, Al-Adhami et al. Al-Adhami et al. [2019] explored the feasibility of Building Information Modeling (BIM) based XR technology for quality control on real construction sites. Johansson and Roupé Johansson and Roupé [2022] found that multi-user VR in a construction context can enhance design and constructability review, sequencing, and job planning. Khairadeen Ali et al. Khairadeen Ali et al. [2021] proposed iVR, a near real-time construction work inspection system that integrates 3D scanning, extended reality, and visual programming to facilitate interactive onsite inspection for indoor activities and provide numeric data. Azhar et al. Azhar et al. [2018] investigated the use of XR in building construction courses to enhance the learning experience and engage students in active learning. Bosché et al. Bosché et al. [2016] developed a mixed reality system for training construction trade workers, which provides realistic and challenging site conditions while mitigating occupational health and safety risks. Balali et al. Balali et al. [2020] proposed a VR-based framework for selecting construction interior finish materials, which incorporates visual aesthetics and cost considerations to assist stakeholders in making informed decisions and managing change orders. Other applications of XR include management Zhao et al. [2023], building operation and maintenance Casini [2022], and visualization and communication Woodward et al. [2010].

Aside from academia, industry has demonstrated real-world applications of XR technology in construction. For example, Industrial Training International offers a VR crane simulator designed for practical training purposes [1].

Unlike previous work, Col-Con is designed as a collaborative construction site simulation testbed, allowing for the implementation of various construction tasks. Col-Con offers researchers a foundational immersive construction environment, including construction vehicles with synchronized events and animations, as well as flexible, pluggable, and configurable features. Researchers can easily set up simulations that support two users. Additionally, we have implemented a realistic pipe installation task on Col-Con and incorporated futuristic human-robot and human-AI interactions.

### 2.2 Collaborative Virtual Environments

Collaborative Virtual Environments (CVEs) are distributed virtual reality systems that provide graphically realized, potentially infinite digital landscapes Churchill et al. [2012]. CVEs are shared by participants across a computer network, where they are given graphical embodiments and can interact with the virtual world's contents and communicate with one another using different media Benford et al. [2001]. A well-known example of a CVE is the Cave Automatic Virtual Environment (CAVE), a projection-based virtual reality display Cruz-Neira et al. [1992]. CAVE is widely used for educational purposes, as demonstrated in De Back et al. [2020], de Back et al. [2021].

With advancements in hardware and software, headset-based CVEs are gaining increased attention. For example, Greenhalgh et al. Greenhalgh and Benford [1995] introduced MASSIVE, a virtual reality teleconferencing system that enables multiple users to communicate through a combination of audio, graphics, and text media over local and wide area networks. Tseng et al. Tseng et al. [2020] explored the effects of a 3D vocabulary learning program on English as a Foreign Language (EFL) young learners' vocabulary acquisition, focusing on learner autonomy and collaboration. Similar CVE-based learning research has been conducted Herrera-Pavo [2021], Jovanović and Milosavljević [2022], Matee et al. [2023]. Lee et al. Lee et al. [2020] developed the Free-roaming Immersive Environment to Support Team-based Analysis (FIESTA), which allows users to position authoring interfaces and visualization artifacts freely within the virtual environment, either on virtual surfaces or suspended in the interaction space. Prabhakaran et al. Prabhakaran et al. [2022] proposed COFFEE, a collaborative virtual environment for furniture, fixtures, and equipment, enabling concurrent multi-user interaction, communication, and collaboration during the design appraisal of interior furnishings. He et al. He et al. [2020] introduced CollaboVR, a reconfigurable framework for both co-located and geographically dispersed multi-user communication in VR, combining animated sketching, collaborative scene editing, and real-time multi-user communication.

Additionally, researchers have proposed using CVEs for various purposes, such as learning Lim et al. [2023], Suzuki et al. [2020], Scavarelli et al. [2021], Mulders et al. [2020], entertainment Oriti et al. [2023], surgery planning Chheang et al. [2021], design Chowdhury and Schnabel [2020], Kalantari and Neo [2020], Tea et al. [2022], healthcare Taylor et al. [2020], and human-robot collaboration Malik et al. [2020].

---

[1] https://www.iti.com/simulations/vr-crane-sim

Different from previous works, Col-Con is a virtual reality simulation testbed specifically designed for construction sites. Furthermore, Col-Con is intended for researchers rather than end users. It serves as a foundational infrastructure for collaboration in construction simulation, allowing researchers to conduct a range of studies, such as construction co-training, collaborative behaviors analysis, and shared situational awareness.

### 2.3 Interaction Techniques for Multi-user VR

Multi-user VR applications are gaining research attention, with advancements in user interface (UI) and interaction techniques. For example, Yuill and Rogers Yuill and Rogers [2012] suggested that successful interfaces should ensure high awareness of others' actions and intentions, provide substantial control over the interface, and offer ample background information. Yassien et al. Yassien et al. [2020] highlighted that asymmetric interactions are under-explored and emphasized that self-embodiment and non-verbal cues are crucial for social presence in social VR applications. Van Damme et al. Van Damme et al. [2023] introduced a networked, distributed multi-user VR system with environment synchronization over low-bandwidth network connections. Yuan et al. Yuan et al. [2023] presented MEinVR, a multimodal interaction technique for exploring 3D molecular data in virtual reality, which combines controller and voice input for data manipulation in immersive environments. Gong et al. Gong et al. [2020] suggested that the quality of the VR environment is critical to the user experience. For more VR interaction techniques in VR, please refer to a recent review Spittle et al. [2022].

From an industry perspective, mature commercial and open-source multi-user frameworks, such as Photon Engine, Unity Multiplayer Networking, and FishNet, are available. Some of these frameworks provide integrated voice communication. Historically, VR applications relied on inverse kinematics (IK) for avatar animation. However, newer frameworks, such as the Meta Movement SDK and Meta Avatar SDK, now offer more advanced solutions for avatar animation and interaction.

Building on previous work, we have developed Col-Con as a multi-user collaborative VR testbed. It utilizes mature multi-user frameworks to support synchronization of transformations, animations, sounds, and interactions, and facilitates real-time voice communication. Additionally, Col-Con leverages the Meta Movement SDK to achieve full-body tracking.

## 3 Overview of Col-Con

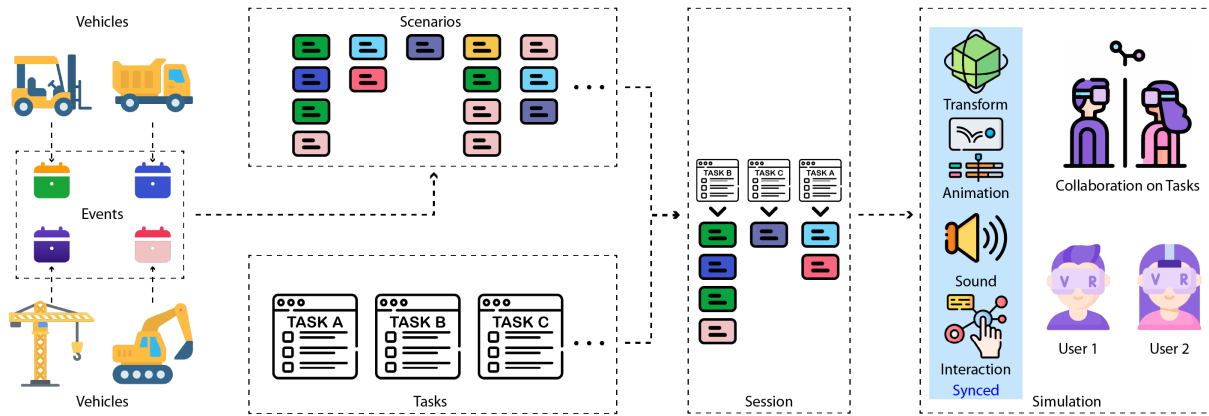

Figure 1: Overview of Col-Con. The platform features construction vehicles positioned within an immersive construction site. Each vehicle operates independently from the environment and other vehicles and is associated with various events. Scenarios are constructed using these events, while construction tasks are independent of the scenarios; thus, scenarios serve as the context in which tasks are executed. A session consists of a combination of scenarios and tasks. Once users join a session, the simulation is prepared to begin. During the simulation, transformations, animations, sounds, and interactions are synchronized as the two users collaborate on tasks.

Fig. 1 shows the overview of Col-Con. It follows a bottom-up architecture. Vehicles serve as the foundation of the simulation, with each vehicle linked to a series of events. These events collectively form the scenario. When combined with the task, the scenario helps build a session, which consists of a series of scenarios and tasks. The session, along with synchronized transformations, animations, sounds, and interactions creates a simulation environment where participants can join and complete tasks collaboratively.

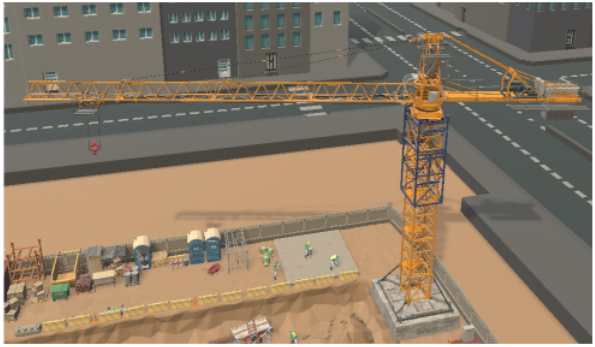
(a) Crane

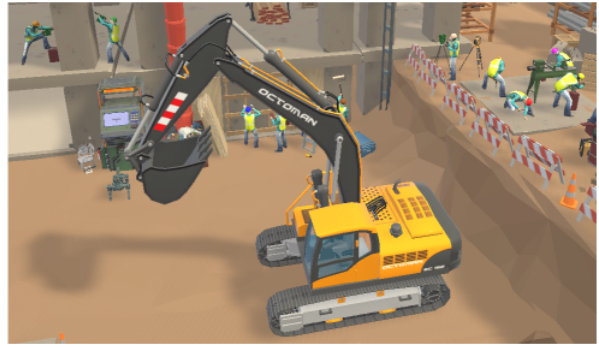
(b) Excavator

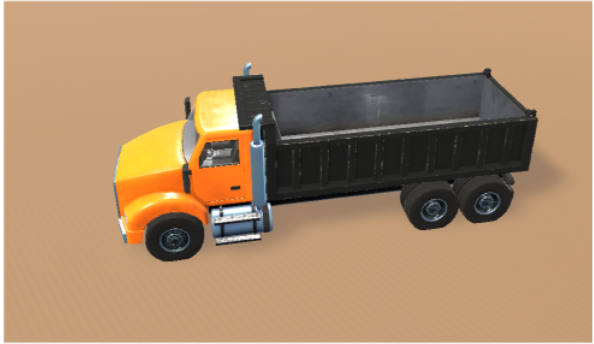
(c) Truck

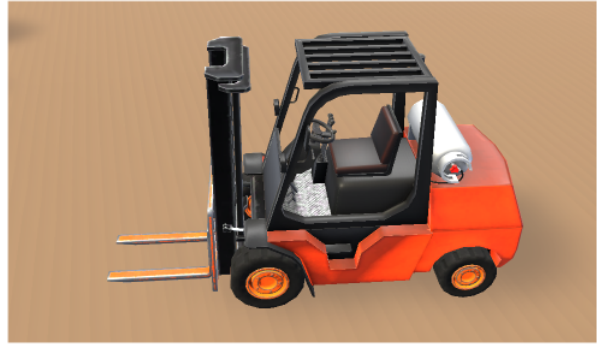
(d) Forklift

Figure 2: Example construction vehicles on Col-Con.

## 3.1 Definitions

The terms are defined as follows. **Vehicles** refer to construction machines. Each vehicle is associated with a series of **events**, which refer to operations or events. Events are categorized as either *normal* or *accident*, depending on whether they lead to a potential hazard. **Tasks** denote construction activities within this environment that participants are required to complete. A **scenario** comprises a series of events. A **session** includes both scenarios and tasks. A **simulation** involves two participants joining a session and collaborating on tasks with synchronized transformations, animations, sounds, interactions, and voice communication.

## 3.2 Configurable Modules

As a fundamental component of Col-Con, some construction vehicles are illustrated in Fig. 2. To facilitate researchers, components such as events, scenarios, and tasks are designed to be modular. We have employed YAML files to achieve this configurability.

The following describes segments from the Crane configuration file. It is a YAML file and is self-explanatory. **Name** represents the configuration file name. **Desc** provides a description of the file. **GameObject** refers to the game object associated with this configuration file. **Events** includes all events related to this vehicle. **Normals** refers to events that do not pose hazards to users, while **accidents** refers to events that may pose hazards to users. Each event has several attributes. **Id** is the unique identifier for the event. **Condition** defines the type of event, either normal or accident. **Desc** is the description of the event. **Warning** contains the warning content of this event, which is converted to audio and played when the event is triggered.

Reflection is used to automatically associate the event implementation with the configuration file. For example, in this configuration file, the event **normals: (id: 1)** is implemented in the method named *Crane_normals_1* within the *Crane* class.

```
1 name: "Crane"
2 desc: "Crane-related events"
```

```
3  gameObject: "Crane"
4  events:
5    normals:
6      - id: 1
7        condition: "Normal"
8        desc: "A load is passing overhead."
9        warning: "Warning: A cargo is passing overhead."
10     - id: 2
11       condition: "Normal"
12       desc: "A hook (without a load) is passing overhead in the opposite direction
           ."
13   accidents:
14     - id: 1
15       condition: "Accident"
16       desc: "A load with an unpacked pipe is being hoisted and is going to pass
         above players."
17       warning: "Warning: A cargo is going to pass overhead."
18     - id: 2
19       condition: "Accident"
20       desc: "A hook (without a load) is passing overhead in the opposite direction
           ."
```

Listing 1: Crane configuration segments. Please refer to the main text for the explanation.

The configuration file for the scenario follows a similar schema to the crane configuration example. This design allows scenarios to be easily constructed by referencing existing events.

### 3.3 Timeline-Driven Event Engine

We developed a customized timeline-driven event engine to manage the sequential triggering of events. This approach provides researchers with the flexibility to specify the timing and content of events by simply creating a scenario configuration file. Additionally, events are synchronized to ensure consistency. For an example of the timeline, please refer to the supplementary material.

### 3.4 Basic Menu System

We implemented a simple menu for Col-Con that always remains in the user's field of view (FOV). This menu is configurable via a configuration file. For illustration purposes, two menu items, *Supervisor* and *Safety Manager*, are included in Fig. 3. Please refer to the supplementary material for more details. Note that the top four menu items in Fig. 3(b) are related to the pipe-installation task, which will be detailed later.

### 3.5 Implementation

We implemented Col-Con on a Windows 11 PC equipped with an NVIDIA RTX 3070 GPU using Unity 2020.3.48. The Meta Quest Pro was employed for both development and user study. We used established multi-user networking frameworks: Photon Fusion for hosting, synchronization, and RPC communications, and Photon Voice for real-time voice communication. The Oculus Interaction SDK was used for basic interactions, such as distance grabbing to allow interaction with virtual objects from a distance. Additionally, Meta Movement SDKwas used for full-body tracking, and FishNet was utilized to synchronize full-body tracking data.

## 4 Example on Col-Con: Pipe Installation

Col-Con is a virtual reality simulation testbed for exploring collaborative behaviors in construction. To showcase its capabilities, we implemented a simulated pipe installation construction task. In this task, two participants are required to install pipes on a wall according to the provided instructions.

### 4.1 3D Models

The basic materials for pipe installation primarily involve pipes, clamps, glue, and connectors. Pipes may vary in diameter. Based on these requirements, we created the 3D models. Our supplementary material shows the model appearances.

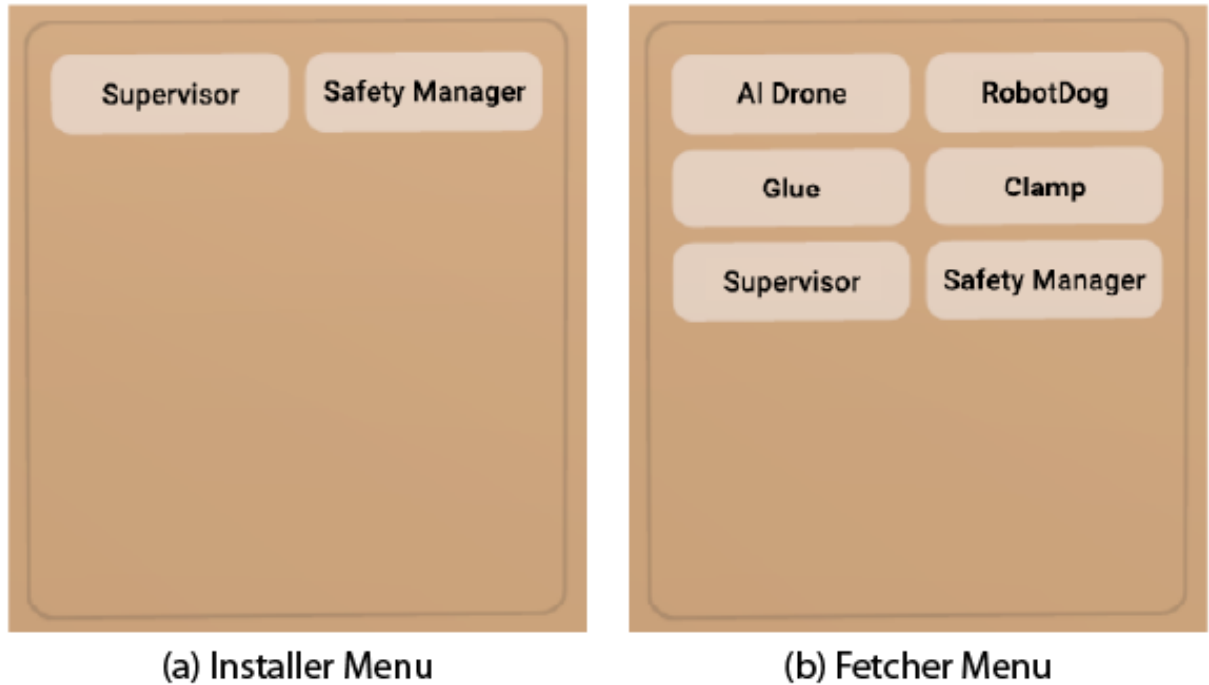

Figure 3: The user menu for the pipe installation task. This menu remains in front of the user's FOV. The Installer and the Fetcher share two menu items: *Supervisor* and *Safety Manager*, derived from the Col-Con. The Fetcher has four unique menu items: *AI Drone*, *RobotDog*, *Glue*, and *Clamp*.

### 4.2 Procedural Pipe Generation

In the construction industry, pipes exhibit a variety of characteristics, including types, diameters, and colors. To accommodate this diversity, we employ procedural techniques to generate different pipe models. In our setup, pipes are specified by four attributes: **type** (sewage, water, gas, and electricity), **color** (magenta, green, blue, and yellow), **diameter** (1, 2, 3, and 4 inches), and **angle** (0°, 45°, 90°, and 135°). Given that the fundamental shape of the pipe models is similar, we use procedural modeling techniques to render variations. We created models for four diameters and four angles, resulting in 16 distinct pipe models without material or texture. To accommodate variations in pipe length, each model, except for the straight pipe, consists of three segments. As illustrated in Fig. 8(a), the green and blue segments can be scaled to adjust the length of the pipes. The type and color attributes are applied through materials during runtime according to specific requirements. Note that the sizes of the pipes are not to scale with real-world dimensions; they are intended for demonstration purposes.

### 4.3 Pipe and Clamp Interactions

To fix the pipe on the wall, clamps are necessary. We use blue hints to mark the areas that require clamping, as shown in Fig. 6(b). The holding point of the pipe is positioned in the middle, which can make it difficult for participants to reach the blue region and place a clamp if the pipe is too long. To address this issue, we introduce an interaction mechanism. Participants can move the pipe left or right by pushing the controller joystick left (Fig. 4(c)) or right (Fig. 4(b)), and can reset by pressing the controller joystick (Fig. 4(a)).

Due to the interaction SDK being utilized, when objects such as pipes and clamps are grabbed, they follow the movement of the controller. This can result in the objects passing through the wall, which diverges from real-world experience. To address this issue, we developed a compensation algorithm and provided haptic feedback to participants. The compensation algorithm projects the pipe's transformation onto the wall surface, ensuring that the object adheres to and remains perpendicular to the surface. The effect of applying the compensation algorithm to the pipe is illustrated in Fig. 6(a) and Fig. 6(b). Similarly, the effect on the clamp is shown in Fig. 7(a) and Fig. 7(b).

When the pipe is nearly horizontal or vertical after compensation, it will be snapped to precisely align horizontally or vertically, accompanied by a long vibration to indicate its orientation. Similarly, after compensating the clamp, a short vibration will signal the participant that they can release the clamp.

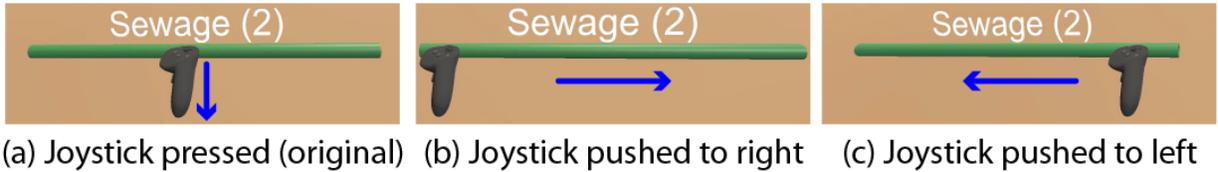

Figure 4: Pipe holding point control. (a) By default, the pipe is held in the middle when pipe is grabbed or when the joystick is pressed to reset; (b) When the joystick is pushed to the right, the pipe moves to the right and the holding point is on the left, and (c) vice versa.

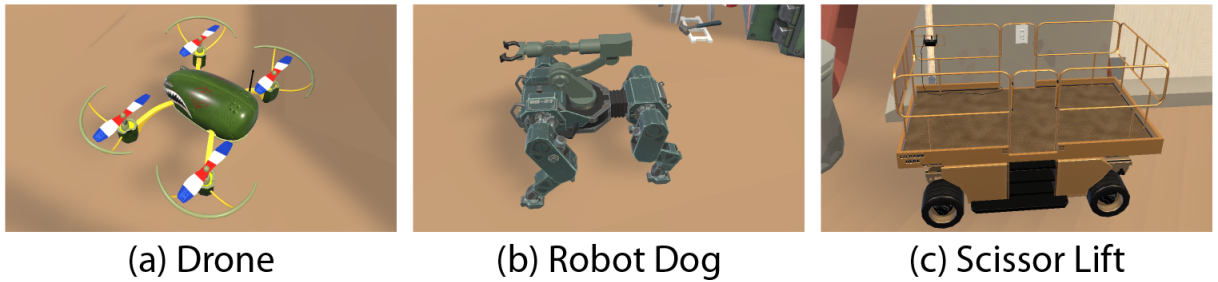

Figure 5: Assistive machines for pipe installation. (a) is a drone for delivering pipes; (b) is a robot dog for carrying pipes for cutting and/or delivering connectors; and (c) is a scissor lift to help with installing pipes that are out of reach.

### 4.4 Arbitrary Pipe-Pipe Connection

Simulating physics-based interactions in VR presents significant challenges, and pipe connecting is one such complex problem. We address this challenge with a novel combined approach from both the model and algorithmic perspectives. Each pipe (except for the $0°$ pipe) consists of three segments as shown in Fig. 8(a). The $x$ axes of the two ends (blue and green segments) are designed always to point outward. By using the transformation and the $x$ direction of one end of a pipe, we can determine the transformation for the connecting end of the other pipe. This allows us to calculate the complete transformation for the other pipe. A container is created to comprise two pipes, resulting in a connected assembly. Additional pipes can also connect to the container.

Given that approximately 90% of people are right-handed De Kovel et al. [2019], we establish a convention that when connecting two pipes, the pipe held by the right controller (right pipe in short) will move towards the pipe held by the left controller (left pipe in short). Fig. 8 illustrates how the connection process works. The green and red axes shown are associated with the connecting end of the left pipe. For instance, the axes in Fig. 8(b) are associated with the green end of the $0°$ pipe, while the axes in Fig. 8(c) are associated with the blue end of the $90°$ pipe. The connecting process starts with a fixed $0°$ pipe and a connecting $90°$ pipe as shown in Fig. 8(b), followed by connections with a $135°$ pipe and a $45°$ pipe.

The connector is treated as a special $90°$ pipe with three fixed-size segments. Consequently, connecting between a pipe and a connector follows the same logic as pipe-to-pipe connecting.

### 4.5 Assistive Machines

To enhance the pipe installation task and introduce futuristic elements, we designed assistive machines as shown in Fig. 5. The drone (Fig. 5(a)) assists with delivering pipes. It is activated by clicking the *AI Drone* button in Fig. 3(b) and clicking the *Confirm* button in Fig. 9. The robot dog (Fig. 5(b)) aids in carrying pipes for cutting and/or delivering connectors. It is triggered by clicking the *RobotDog* button in Fig. 3(b) and clicking the *Confirm* button in Fig. 10.

The scissor lift (Fig. 5(c)) assists participants in performing out-of-reach tasks. Participants can enter the lift when they approach it and the "Press X to Enter" hint appears. The hint changes to "Press X to Exit" once they enter. The left-hand joystick is used to move the lift left, right, up, and down.

### 4.6 Pipe Installation Procedure

Fig. 11 shows a pipe installation example with the following steps:

(i) Grab a pipe and a connector (Fig. 11(a));

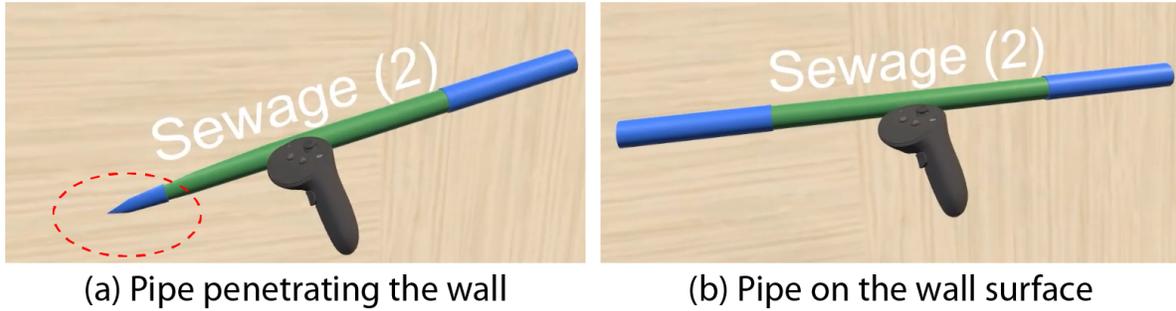

(a) Pipe penetrating the wall     (b) Pipe on the wall surface

Figure 6: Compensation for holding a pipe. Due to the interaction SDK in use, the held pipe follows the controller's movement, resulting in an unrealistic wall penetration artifact. To address this issue, we developed a compensation algorithm that adheres the pipe to the wall surface. (a) and (b) show the effect before and after applying the compensation algorithm respectively. The blue regions indicate where clamps should be placed. These blue regions appear when the pipe touches the wall and disappear either when the pipe is removed from the wall, or when a correctly-sized clamp is placed on the blue region and the pipe is fixed to the wall. Note that when the pipe is nearly horizontal or vertical after compensation, it will be snapped to be perfectly horizontal or vertical, followed by a long vibration signal to inform the user about the snapping.

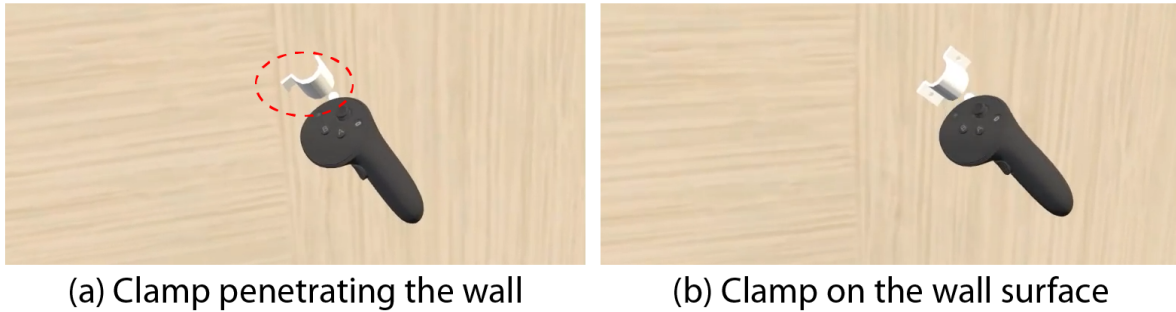

(a) Clamp penetrating the wall     (b) Clamp on the wall surface

Figure 7: Compensation for holding a clamp. A similar wall penetration artifact appears when a clamp is held and moved with the controller. To address this issue, we developed a separate compensation algorithm. (a) and (b) show the effect before and after applying this algorithm, respectively. Note that a short vibration will be triggered when the clamp has been compensated, which informs the participant that he can release the clamp.

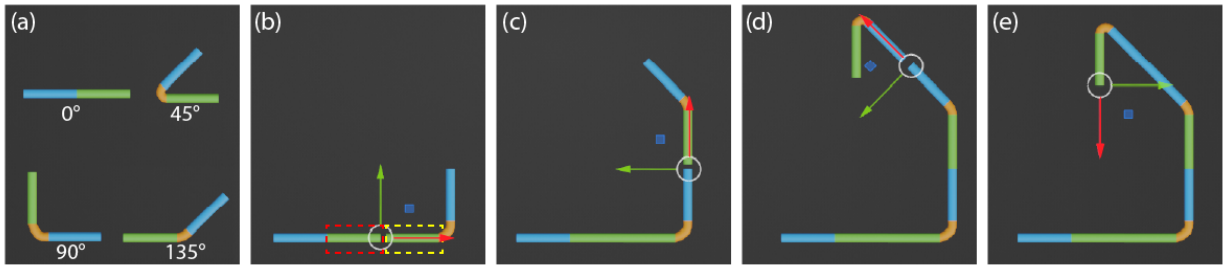

Figure 8: Illustration of the arbitrary pipe connecting algorithm. Each pipe is divided into two or three segments colored blue, green, and orange for illustrative purposes; these colors do not appear in the actual models. (a) shows pipe models with different angles: 0°, 45°, 90°, and 135°. (b) demonstrates how to connect a 0° pipe with a 90° pipe. The 0° pipe is fixed, and the extending direction (depicted by the red arrow) of the contact part (boxed in red) is calculated. The contact part of the 90° pipe (boxed in yellow) will move along with the entire 90° pipe, resulting in a seamless connection. The same rule applies to connecting (c) a 135° pipe and (d) a 45° pipe. (e) This process can be repeated to connect additional pipes at the end.

(ii) Place a clamp after experiencing a long vibration (indicating the pipe is horizontal) and release the clamp after a short vibration (Fig. 11(b));

(iii) Place a clamp on the other end of the pipe and release it after a short vibration (Fig. 11(c));

(iv) Apply glue (indicated by the purple hint showing its glued status) (Fig. 11(d));

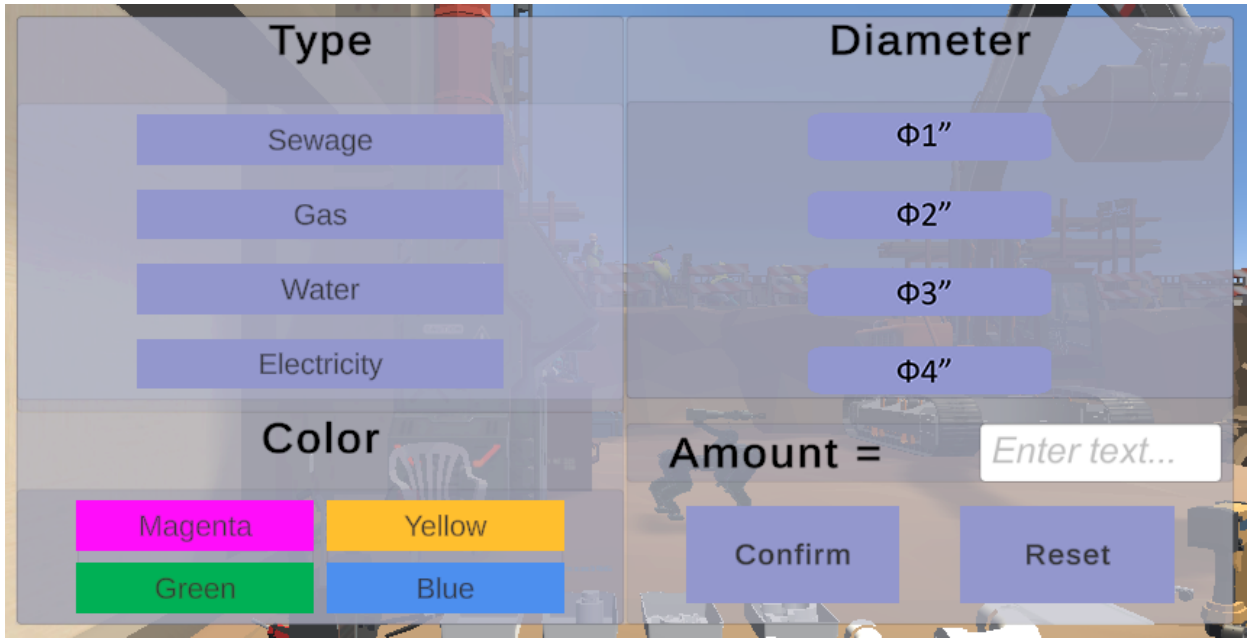

Figure 9: The Drone UI for ordering pipes: This interface appears after the Fecther presses the *AI Drone* menu button (Fig. 3(b)).

(v) Attach a connector of the same size to the glued end (Fig. 11(e));

(vi) Apply glue to the connector (Fig. 11(f));

(vii) Position another pipe on top of the connector (Fig. 11(g));

(viii) Place a clamp on the blue region and release it after a short vibration (Fig. 11(h)).

### 4.7 Task-Specific Menu System

To simulate the pipe installation process, we define two roles: *Installer* and *Fetcher*. They have distinct menus as shown in Fig. 3. The Installer is responsible for installing pipes on the wall and has interactions including grabbing pipes, clamps, and connectors; clamping; gluing; connecting; and operating the scissor lift. The Fetcher supports the installation process and has interactions such as finding and moving pipes from storage, commanding the drone to deliver pipes (by pressing the **AI Drone** button in Fig. 3(b), which opens the window shown in Fig. 9), commanding the robot dog to cut pipes and/or deliver connectors (by pressing the **RobotDog** button in Fig. 3(b), which opens the window shown in Fig. 10), and refilling glue (by pressing the **Glue** button) and clamps (by pressing the **Clamp** button).

### 4.8 Configurable Collaborative Tasks

As aforementioned, pipe specifications involve four types of information: type, color, diameter, and angle. For the pipe installation task, we use the type, color, and diameter specifications and add an additional specification: length. To simplify, only straight (0°) pipes are used. The Installer and the Fetcher are given partial information to encourage collaboration. For instance, if the Installer receives information about color and length, the Fetcher will be provided with information about size and type. This division requires them to exchange information to complete the task. Following the design principles of Col-Con and ensuring flexibility and configurability, we use YAML files for configuration. For more details, please refer to the supplementary material.

### 4.9 Implementation

We implemented the pipe installation task within the same environment as the Col-Con testbed, reusing fundamental components that Col-Con provides. Additionally, we configured two scenarios to accompany the training and pipe installation tasks. These scenarios include several events that are triggered in sequence at specified times once a session begins.

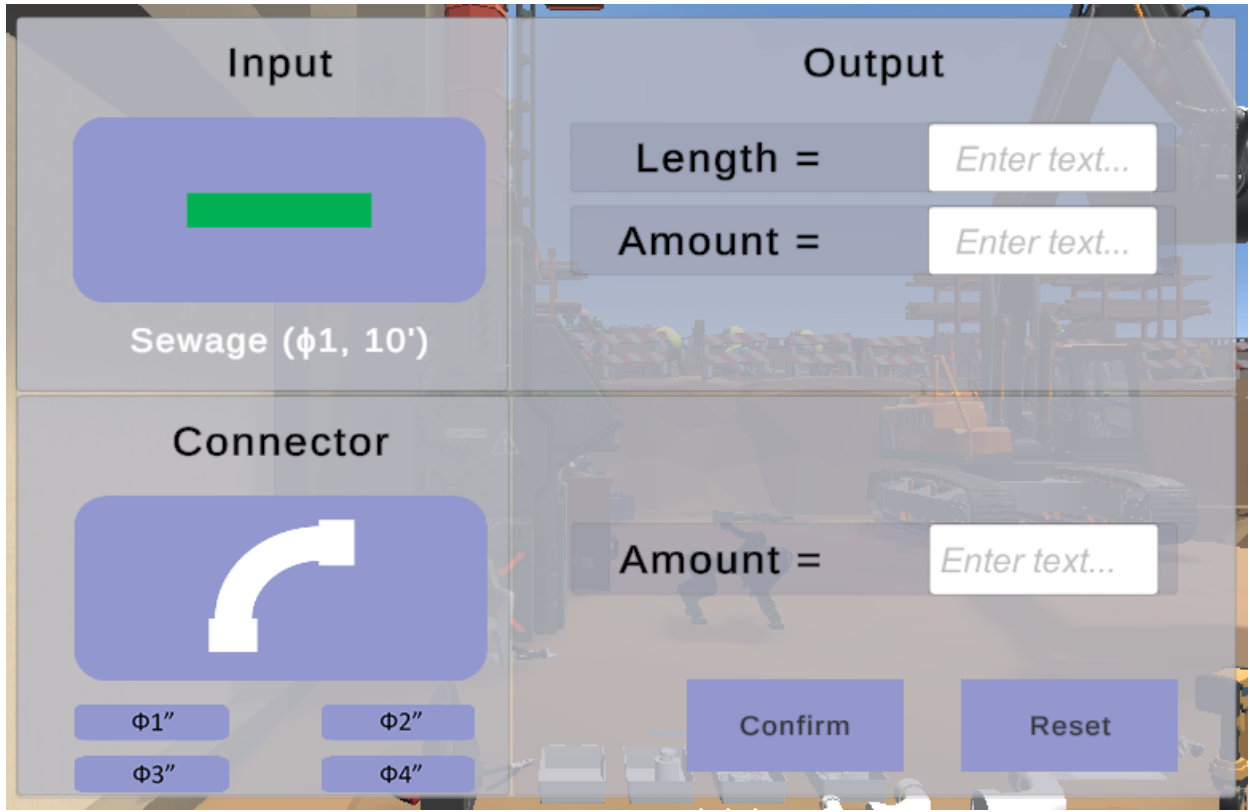

Figure 10: The robot dog UI for cutting pipes and delivering connectors: This interface appears after the Fetcher clicks the *RobotDog* menu button (Fig. 3(b)).

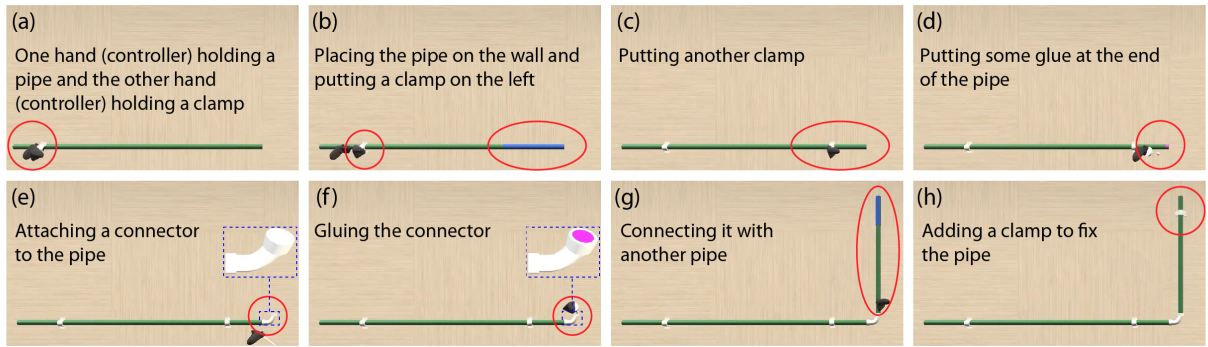

Figure 11: A step-by-step example of pipe installation. (a) shows one controller holding a pipe and the other holding a clamp. (b) demonstrates using the clamp to secure one end of the pipe, with the right blue region indicating the clamping area. (c) illustrates using another clamp to fix the pipe. (d) depicts applying glue to the right end of the pipe. (e) shows placing a connector at the right end of the pipe. (f) depicts adding glue to the connector. (g) shows connecting another pipe to the connector. (h) presents the final setup after placing a clamp on the top end.

## 5 User Study

In a user study, we evaluated our proposed Col-Con testbed alongside the pipe installation task implementation. Specifically, we let participants evaluate different aspects such as usability, immersiveness, motion sickness, and team collaboration supported by having them go through collaborative pipe installation tasks on Col-Con. The user study was conducted on two Windows 11 PCs: one equipped with an NVIDIA GTX 1070 GPU and the other with an NVIDIA RTX 2070 GPU. To minimize latency, the Meta Quest Pro headset was connected to the PCs via a cable during the study.

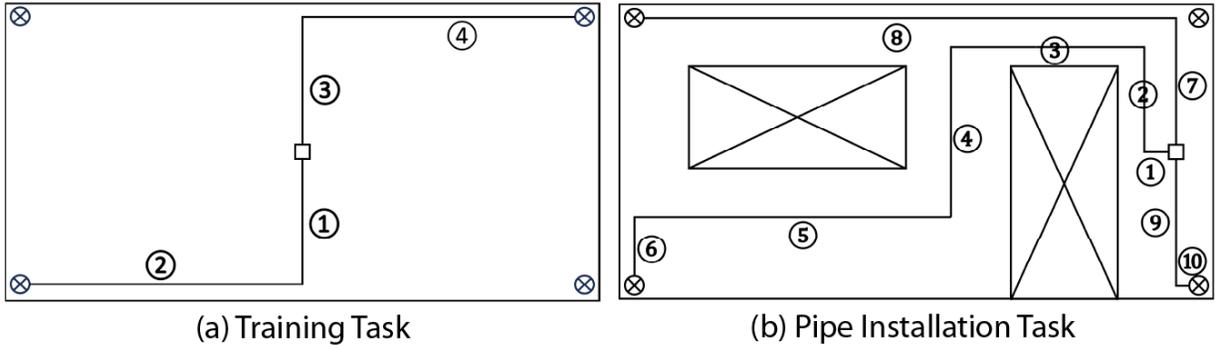

Figure 12: Target pipe layouts. The symbol $\otimes$ represents the endpoint; the symbol $\square$ indicates the box for disconnected pipes; and circled numbers denote pipe segments. Each perpendicular junction between two pipe segments requires a connector. (a) and (b) depict the layouts for the training and pipe installation tasks, respectively. The corresponding pipe specifications are provided in Table 1.

## 5.1 Questionnaire

We designed a questionnaire to gather participants' responses, covering six aspects:

- **Demographics**: Collecting information including genders, ages, and VR experience.
- **Usability**: Evaluated using the System Usability Scale (SUS) Brooke et al. [1996], which includes 10 five-point questions.
- **Immersiveness**: Assessed through the Igroup Presence Questionnaire (IPQ) Usoh et al. [2000], consisting of 14 seven-point questions.
- **Motion Sickness**: Measured with the Simulator Sickness Questionnaire (SSQ) Kennedy et al. [1993], featuring 16 three-point questions.
- **Team Collaboration**: Evaluated using a modified Group Cohesion Questionnaire Khatamian Far [2020]. We included four task cohesion-related five-point questions: (1) *I feel that we all have a common understanding of the task*; (2) *I feel that group members are very united to achieve our goal on this project*; (3) *I feel that individuals associated with my team have a desire to perform well*; and (4) *I feel that my team is committed to the task*.
- **General Feedback**: Collecting participants' opinions on features they liked or disliked and suggestions for improvement.

## 5.2 Procedure

This user study was conducted in a university laboratory with two separate rooms connected by a door, which remained closed during the study. The experimenter first provided an overview of the user study to the two participants. They then watched two instructional videos: the first from the perspective of the Installer and the second from the perspective of the Fetcher. After viewing the videos, the participants decided who would assume each role.

Two sessions were configured on the Col-Con testbed: a training task without events and a pipe installation task with 10 sequential events. During the pipe installation task, events occurred in a predefined order while the participants worked together to install pipes. Voice communication between the participants was facilitated via the Quest Pro headsets.

Following the role assignment, participants completed the training task to learn basic interactions, including grabbing, clamping, connecting, ordering, and cutting pipes. The experimenter provided guidance as needed.

After the training task, participants proceeded to the pipe installation task, where the interactions remained the same but the tasks differed. Participants were asked to complete a questionnaire on the PC after completing the pipe installation task.

## 5.3 Participants

The university's Institutional Review Board approved this study. We recruited participants via emails and social platforms. A total of 28 participants (self-reported: 11 males and 17 females, aged 21-38, M = 26.36, SD = 4.26) were recruited. 16 participants reported no VR experience, while 12 participants reported having years of VR experience (M = 2.75, SD = 1.66). 7 participants had VR gaming experience; 7 participants experienced VR in research/course

Col-Con: A Virtual Reality Simulation Testbed for Exploring Collaborative Behaviors in Construction

Table 1: Task specifications for pipe layouts in Fig. 12. Note that the Installer and the Fetcher are provided with different types of information to prompt collaboration between them.

| Session | Training Task | | | | Pipe Installation Task | | | |
|---|---|---|---|---|---|---|---|---|
| Role | Installer | | Fetcher | | Installer | | Fetcher | |
| Segment | Color | Type | Size | Length | Color | Length | Size | Type |
| 1 | green | sewage | 1 | 4.2 | green | gas | 1 | 1 |
| 2 | green | sewage | 1 | 9.8 | blue | gas | 1 | 3.5 |
| 3 | blue | gas | 3 | 3.8 | green | gas | 1 | 7.5 |
| 4 | blue | gas | 3 | 9.8 | blue | gas | 1 | 6 |
| 5 | - | - | - | - | green | gas | 1 | 10 |
| 6 | - | - | - | - | blue | gas | 1 | 2.5 |
| 7 | - | - | - | - | yellow | electricity | 4 | 3.5 |
| 8 | - | - | - | - | magenta | electricity | 4 | 18.5 |
| 9 | - | - | - | - | magenta | water | 2 | 4 |
| 10 | - | - | - | - | blue | water | 2 | 0.5 |

projects; 1 participant experienced VR in attractions; and 1 participant tried VR in movies. Each user study session comprised a team of two participants who collaborated to complete tasks.

### 5.4 Results

In the user study, we found that users could collaborate effectively. Additionally, users felt that the simulation and the pipe installation task were realistic and engaging. Fig. 13 shows an example of the progress made by participants during the pipe installation task.

#### 5.4.1 Usability

We calculated the SUS score based on the process devised by John Brooke Brooke et al. [1996]. The overall average of SUS scores is 77.59 with an SD of 13.67. According to the approximations of the adjectives specified by Bangor et al. Bangor et al. [2009], 77.59 is regarded as "good". Specifically, the number of "excellent" ratings is 9 (32%), the number of "good" ratings is 9 (32%), and the number of "ok" ratings is 10 (36%).

We performed a one-sample t-test, hypothesizing a significant difference between the SUS scores and the good standard (72.5), specifically that the SUS scores mean was greater than the good standard. The one-tailed analysis result is t = 1.97 and $p < 0.05$, supporting the hypothesis.

Moreover, we found a statistically significant difference between the Installer and the Fetcher (t = -2.73, $p < 0.05$). This is reasonable, as the Installer's task requires more interactions compared to the Fetcher's task.

#### 5.4.2 Presence

Per Usoh et al. Usoh et al. [2000], we first calculated the spatial presence, involvement, and experienced realism. Then, we calculated their characteristic scores. Table 2 shows the results. As the IPQ questionnaire uses a 7-Points-Likert scale, we can conclude that the scores for spatial presence, involvement, and experienced realism are moderate. This means that Col-Con makes the user feel present, engaged, and realistic to some extent.

We did not find statistically significant differences between the Installer and the Fetcher in all three aspects: Spatial Presence (t = 1.39, p = 0.19), Involvement (t = 0.77, p = 0.46), and Experienced Realism (t = 0.50, p = 0.63).

These results indicate that the presence experienced in Col-Con is role-independent, meaning participants within the group have the same immersive experience.

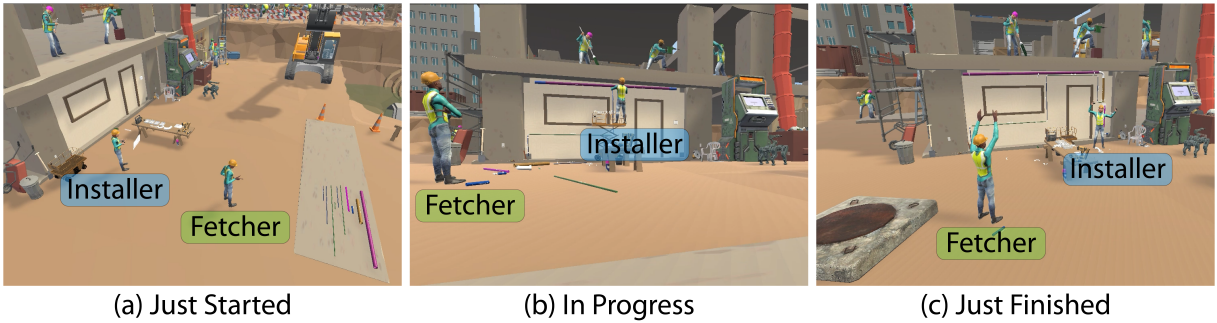

Figure 13: An example of the pipe installation task observed in the user study. In (a), the task had just started; the Fetcher was examining the instruction sheet, while the Installer was checking the storage area where some pipes were placed. In (b), the installation was in progress; the Installer was working on the installation while the Fetcher observed. In (c), the installation was complete, and the users waved to each other to celebrate.

### 5.4.3 Motion Sickness

Per Kennedy et al. Kennedy et al. [1993], we calculated the mean and standard deviation of Nausea, Oculomotor, Disorientation, and Total Score from SSQ responses. Table 3 shows the results. Compared to a previous study that used the same SSQ questionnaire to evaluate motion sickness Liu et al. [2023], the total score of Col-Con is 13.22, which is lower than the 19.77 reported in their work. Furthermore, the average scores for Nausea, Oculomotor, and Disorientation in Col-Con are lower, indicating that the level of sickness caused by Col-Con is acceptable and may not significantly impact participants' comfort.

We did not find statistically significant differences between the Installer and the Fetcher in all four aspects: Nausea (t = 0.45, p = 0.66), Oculomotor (t = 0.19, p = 0.85), Disorientation (t = 0.84, p = 0.42), and Total Score (t = 0.53, p = 0.60).

### 5.4.4 Collaboration

Based on all 28 participants' ratings, the overall task cohesion ratings are as follows: Mean = 4.75, and SD = 0.35. As these four questions use the 5-Points-Likert scale, we conclude that collaboration is high and effective.

We did not find statistically significant differences between the Installer and the Fetcher in collaboration ratings (t = 0.43, p = 0.67), which indicates that both participants within a group have similar perceptions of collaboration.

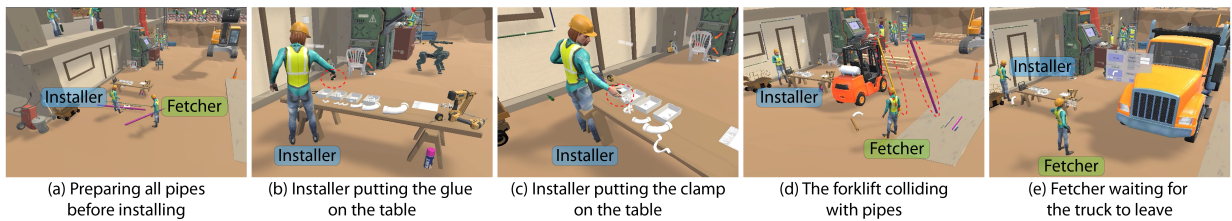

(a) Preparing all pipes before installing  (b) Installer putting the glue on the table  (c) Installer putting the clamp on the table  (d) The forklift colliding with pipes  (e) Fetcher waiting for the truck to leave

Figure 14: Interesting moments captured from the user study. (a) Some participants preferred to prepare all pipes before beginning the installation process. (b) The Installer often placed the glue on the table after use, despite not being instructed to do so. (c) The Installer would place the clamp on the table if it did not fit the pipe properly. (d) A passing forklift collided with pipes prepared by the Fetcher, causing the pipes to bounce. (e) The truck obstructed a portion of the UI, causing the Fetcher to wait for the truck to leave. It is worth noting that the Fetcher could move to the left, triggering the UI to follow and become fully visible.

### 5.4.5 General Feedback

At the end of the questionnaire, we asked the participants four questions and summarized their responses as follows. We use the abbreviation *Group number-Role* to denote the participant within a given group. For example, G1-I and G1-F refer to the Installer and Fetcher in Group 1, respectively.

**Q1: What features do you like?**  Participants expressed their preferences for various features in the following ways. Several participants, including G1-F, G8-F, G10-F, G12-F, and G14-F, complimented the drone and robot dog, with G10-F noting, "The robot dog was a nice idea." The immersive environment was praised by G1-I, G7-I, and G9-F, with G1-I commenting, "The virtual world is an excellent replica of reality." Environmental sound was valued by G2-F and G3-F, while G7-F found the task was realistic.

Table 2: Results of the IPQ questionnaire. Metrics are calculated based on all 28 participants' ratings.

| Metrics | Spatial Presence | Involvement | Experienced Realism |
|---|---|---|---|
| Mean | 4.38 | 4.59 | 4.42 |
| SD | 0.65 | 0.72 | 1.21 |

Table 3: Results of the SSQ questionnaire. Metrics are calculated based on all 28 participants' ratings.

| Metrics | Nausea | Oculomotor | Disorientation | Total Score |
|---|---|---|---|---|
| Mean | 11.93 | 17.60 | 12.43 | 13.22 |
| SD | 17.92 | 19.69 | 16.66 | 15.05 |

Participants G6-F, G6-I, G8-I, G8-F, and G10-I appreciated the pipe interactions. Clear instructions were favored by G4-I and G13-F. The intuitive and realistic interactions were noted positively by G9-I, G9-F, G11-I, and G13-I. Vibration feedback was appreciated by G11-I and G14-I, and the scissor lift feature was liked by G5-I and G10-I. Real-time voice communication was valued by G9-F and G13-F, and G4-I found collaboration in VR beneficial.

Additionally, G4-F enjoyed "learning while playing," while G13-I highlighted their appreciation for "the realistic way of fixing pipes on the wall, including the glue."

**Q2: What features don't you like?** Some participants expressed their dislikes regarding various aspects of the experience. G1-F found it confusing to order multiple different pipes simultaneously, particularly when they were of the same type but different colors. G4-F reported discomfort due to a simulated truck nearly hitting her and suggested incorporating additional safety features. G5-F and G8-F experienced issues with the accuracy of targeting and inputting numbers. Both G8-I and G11-I disliked riding the scissor lift, with G11-I noting that the movement up and down made her feel somewhat dizzy. G6-I and G12-I were dissatisfied with the need to carefully align pipes on the wall sometimes. G9-F mentioned that the process became somewhat disorganized after ordering connectors and clamps multiple times. G10-I commented, "I wasn't able to see the full view when I got too close to the wall." Besides, G4-I stated, "I liked everything."

**Q3: Any suggestions to improve it?** Some participants provided suggestions for improvement. G1-F suggested two participants could have the full information, which might defeat the purpose of a collaborative simulation but could speed up the process. Suggestions for enhancing interactions included improvements to grabbing (G1-I), targeting (G5-F), and putting items back (G9-I). Participants also recommended design revisions, such as adding tips (G7-F), displaying connector sizes (G12-I), reorganizing refilled clamps (G9-F), and ensuring consistent environment sizes (G8-I). G10-F preferred movement using a joystick.

Some participants expressed positive feedback. G4-F commented, "I really had a very great experience!" G5-I described the experience as "great", and G13-I stated, "It was a good experience."

**Q4: Anything else you want to tell us?** Participants expressed their favor in response to the question. G1-F commented, "Good work." G1-I remarked, "It is good to see people coming up with unique solutions for the problems we face in the world. Keep up the excellent work and continue your amazing efforts." G4-F praised the effort with, "Great job! Good luck with your team's future endeavors." G4-I found the VR experience particularly valuable, noting, "It's really interesting to build in VR. As a CS graduate with limited knowledge of the construction field, I learned a lot through this VR experience. Thanks for designing this, and kudos to your team." G9-F described the work as "Nice." G10-I found the game "interesting", while G13-I described it as a "fun experience."

# 6 Discussion

## 6.1 Task Completion Guidance

The only rule communicated to participants was that they should install pipes from left to right as specified in the instruction sheet. Beyond this, they had the freedom to decide on various aspects, such as the installation order, the use of existing pipes, ordering new pipes, and cutting pipes. This level of autonomy led to some interesting observations. For instance, most groups alternated between preparing and installing pipes—preparing a pipe and then installing it. However, some groups preferred to prepare all the pipes before starting the installation process (Fig. 14(a)). Additionally, it was noted that some groups continuously ordered and cut pipes, but did not use those pipes, leaving them in the storage area.

Another interesting observation was that some participants, particularly Installers, placed the glue (Fig. 14(b)) or the clamp (Fig. 14(c)) back on the tabletop, even though this was not specified in the instructions. This behavior suggests that some participants preferred to maintain an organized workspace, while others did not exhibit the same inclination.

### 6.2 Unexpected Situations

We designed events in Col-Con to simulate real-world scenarios that participants might encounter during pipe installation. For instance, if pipes were placed in the path of a truck or forklift, collisions could occur, causing the pipes to be displaced due to physics. In such cases, participants needed to reorder and/or cut pipes again.

For example, Fig. 14(d) illustrates a forklift colliding with pipes, while Fig. 14(e) shows the Fetcher waiting for the truck to clear the area. It is important to note that the user interface (UI) moves with the participants, ensuring that it remains in their view. As a result, the Fetcher has the option to move left to maintain visibility of the UI. Additionally, some participants proactively reminded their partners about incoming vehicles.

### 6.3 Limitations and Future Work

Currently, Col-Con supports two-user collaboration in the construction site. Given that construction sites naturally involve multiple workers, expanding the system to support more users would enhance its realism. The current implementation of Col-Con is based on Photon Fusion, which, due to its design regarding input authority and state authority, limits interaction flexibility between users. For future work, we plan to transition to Unity Multiplayer Networking, as its distributed authority model is better suited for complex multi-user interactions. Additionally, based on feedback from group collaboration questions, Col-Con has proven to be an effective testbed for multi-user collaboration in a simulated construction environment. Moving forward, we aim to explore and incorporate a wider range of collaborative construction tasks.

In the simulated pipe installation task, we defined two distinct roles: Installer and Fetcher. Each role is responsible for different tasks, as reflected in their respective menus and interactions. For example, the Installer cannot access the AI Drone and RobotDog menus, limiting their ability to order or cut pipes. Additionally, information about pipe specifications is distributed between the roles, necessitating communication and information exchange to complete the tasks. As noted by G1-F, *"Two participants could have the full information, which might defeat the purpose of a collaborative simulation but could speed up the process."* This feedback suggests the need to explore a more realistic and balanced setting, where two users can collaborate both naturally and efficiently.

From a user experience perspective, Col-Con currently requires a cable connection between the headset and the PC, meaning it is not a standalone application. This setup limits the range within which users can move and affects the user experience, as they need to manage the risk of the cable becoming tangled. While we attempted to run Col-Con wirelessly using Air Link, the performance did not meet our satisfaction compared to the wired connection. For future work, we plan to enable Col-Con to operate wirelessly without compromising the user experience.

## 7 Conclusion

We propose Col-Con, a virtual reality simulation testbed for exploring collaborative behaviors in construction. Col-Con offers a highly immersive experience, allowing researchers to easily configure scenarios using configuration files. The testbed ensures synchronized transformations, animations, sounds, interactions, and real-time voice communication, enabling users to be fully immersed in the same environment. We have implemented a realistic pipe installation task featuring rich, realistic, and synchronized interactions. We also explored futuristic construction scenarios involving human-robot and human-AI interactions in this pipe-installation task. A user study involving 14 groups (28 participants) was conducted to evaluate both Col-Con and the simulated construction task. The results indicate that Col-Con is considered good for use, with a moderate level of immersion and low, acceptable motion sickness. Additionally, Col-Con is task/role independent, and participants reported similar experiences regarding presence, motion sickness, and collaboration. We envision Col-Con as a multi-user simulation testbed that can facilitate research on virtual reality-based collaborative behaviors exploration in the construction context, such as co-training, behavior analysis, and shared situational awareness.